\documentclass[10pt,aps,reprint,prl,twocolumn,amsmath,amssymb,superscriptaddress,showpacs,floatfix]{revtex4-1}
\usepackage{graphicx}
\usepackage{bm}
\usepackage[utf8]{inputenc}
\usepackage[T1]{fontenc}
\usepackage[english]{babel}
\usepackage{tikz}
\usepackage{grffile}
\usepackage{gensymb}
\usepackage{pdfpages} 
\usepackage{pgffor}
\usepackage[caption = false]{subfig}
\usepackage[export]{adjustbox}

\makeatletter
\AtBeginDocument{\let\LS@rot\@undefined}
\makeatother

\begin{document}
\title{The Raman fingerprint of rhombohedral graphite}

\author{Abderrezak torche}
\email{abderrezak.torche@upmc.fr}
\affiliation{
	IMPMC, CNRS, Universit\'e P. et M. Curie, 4 Place Jussieu,
	75005 Paris, France}

\author{Francesco Mauri}
\email{francesco.mauri@uniroma1.it}
\affiliation{
	Universit\'a di Roma La Sapienza,Piazzale Aldo Moro 5, I-00185 Roma, Italy}

\author{Jean-Christophe Charlier}
\email{jean-christophe.charlier@uclouvain.be}
\affiliation{
	Universit\'e catholique de Louvain , Institute of Condensed Matter and Nanoscience , Chemin des \'etoiles 8, 1348 Louvain-la-Neuve, Belgium}

\author{Matteo Calandra}
\email{matteo.calandra@upmc.fr}
\affiliation{
	IMPMC, CNRS, Universit\'e P. et M. Curie, 4 Place Jussieu,
	75005 Paris, France}
	
\begin{abstract}
	Multi-layer graphene with rhombohedral stacking is a promising carbon phase 
	possibly displaying correlated states like magnetism or superconductivity due to the occurrence
	of a flat surface band at the Fermi level.
        Recently, flakes of thickness up to 17 layers
	were tentatively attributed ABC sequences although the Raman fingerprint
	of rhombohedral multilayer graphene is currently unknown and the 2D resonant Raman spectrum
	of Bernal graphite not understood.
	We provide a first principles description of the 2D Raman peak
	in three and four layers graphene  (all stackings) as well
        as in Bernal, rhombohedral and 
	an alternation of Bernal and rhombohedral graphite. 
	We give practical prescriptions to identify long range sequences of ABC multi-layer graphene. Our work is a prerequisite to experimental 
	non-destructive identification and synthesis of rhombohedral graphite.
\end{abstract}
	
\maketitle

\begin{figure}[t]
	\centering 
	\includegraphics[width=\columnwidth]{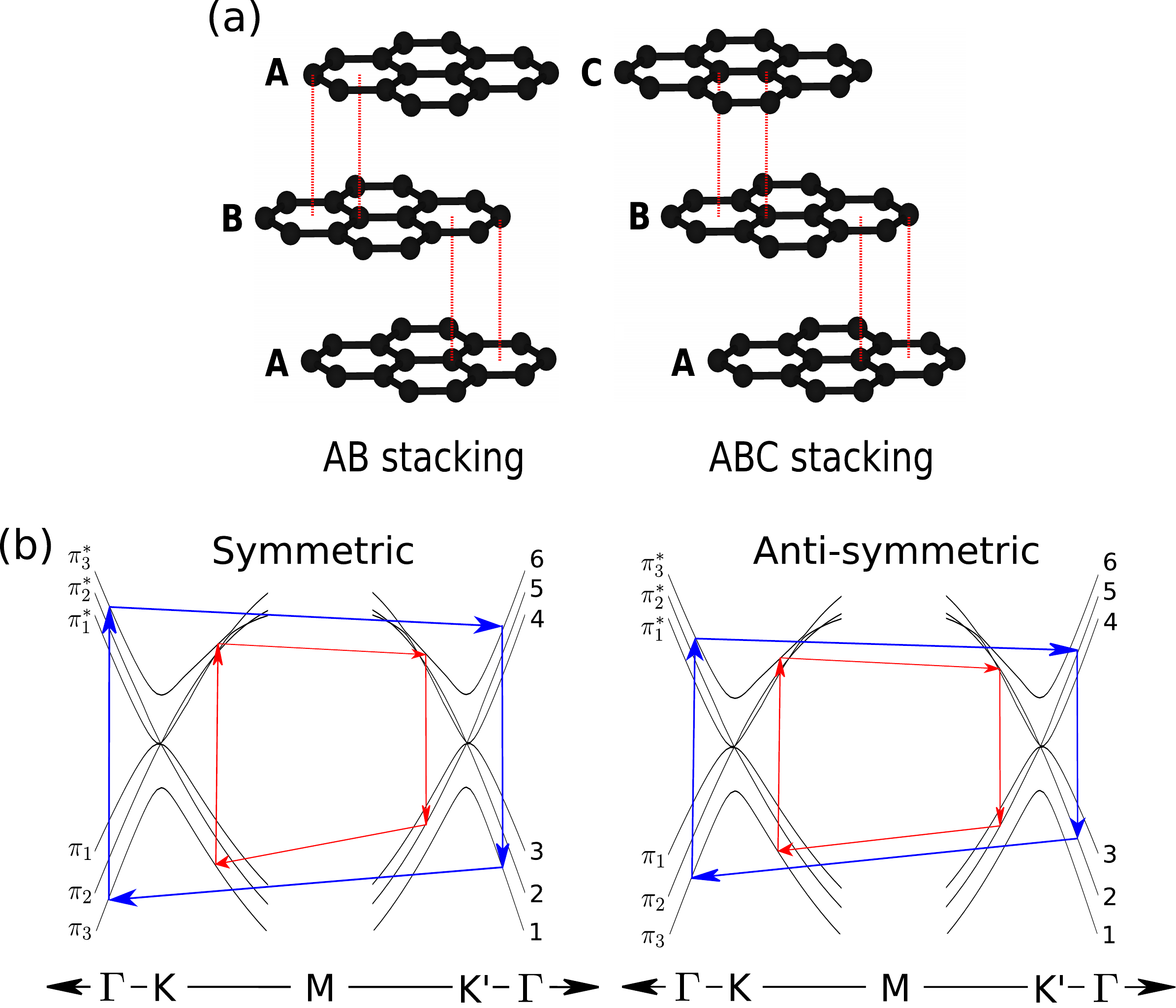}
	\caption{(a) Crystal structure of Bernal (ABA) and rhombohedral (ABC) stacked multilayer graphene. (b) Cartoon of symmetric, asymmetric, inner (red) and outer (blue) double resonant Raman processes in trilayer graphene.}
	\label{fig1}
\end{figure}

Bernal graphite \cite{bernal1924structure} with AB stacked  graphene is the most stable form of graphite.
Recently, however, rhombohedral stacked multi-layers graphene (RMG) with ABC stacking, see Fig. \ref{fig1} (a), attracted an increasing attention as theoretical 
calculations suggest the occurrence of a dispersionless electronic band
(bandwith smaller than 2 meV) at the Fermi level \cite{kopnina2014surface,xiao2011density}. 
This flat band with extremely large effective mass, constitutes a very promising candidate for highly correlated 
states of matter such as  
magnetism \cite{PamukPhysRevB.95.075422} or room-temperature superconductivity \cite{Precker_Esquinazi}. 

As ABC-stacked graphite is metastable  \cite{charlier1994first}, the synthesis of long sequences of ABC graphene layers is a real
challenge. For a random sequence of $N$ graphene layers stacked along
the $c$ axis, a purely statistical argument states that the probability to obtain 
$N$ layers with ABC order is $1/2^{N-1}$. In reality the
probability is even lower as all stackings are not equally probable as
energetics favour the Bernal one with respect to the others.
This explains why three and four layer graphene flakes with
ABC-stacking are 
systematically found \cite{PhysRevLett.104.176404,lui2010imaging,cong2011raman}, while it is highly unprobable
to obtain long range ABC-stacking order. Recently, it has been suggested that pentalayers graphene with 
rhombohedral stacking can be grown epitaxially on 3C-SiC(111) \cite{pierucci2015evidence}. Finally, Henni {\it et al.}  \cite{henni2016rhombohedral}
were able to isolate multilayer graphene flakes with ABC sequences exceeding 17 graphene sheets.
However, while for three and four ABC stacked graphene layers an optical signature exists \cite{PhysRevLett.104.176404,lui2010imaging},
a clear fingerprint of long-range rombohedral order is lacking.

Raman spectroscopy, and in particular the 2D double resonant Raman peak, has proven to be a very 
powerful technique to investigate  structural and physical properties of graphene flakes. It can be used 
to count the number of layers \cite{ferrari2006raman}, detect charged impurities \cite{casiraghi2007raman}, 
measure the strain-induced deformation of the electronic structure \cite{Huang_stress_2D,Mohr_strain}, measure the phonon dispersion 
 \cite{PMayPhysRevB.87.075402, herziger2014two} and many other properties (for a review see  \cite{ferrari2013raman}).
However, despite its crucial importance, the theoretical understanding of the 2D double resonant Raman spectrum has been
obtained only for graphene  \cite{VenezuelaPhysRevB.84.035433,Basko_PhysRevB.78.125418,PopovPhysRevB.87.155425} and 
bilayer graphene \cite{herziger2014two}. Even the basic case of bulk Bernal graphite is not completely understood.
\begin{figure*}[t!]
	\centering 
	\includegraphics[width=0.49\textwidth]{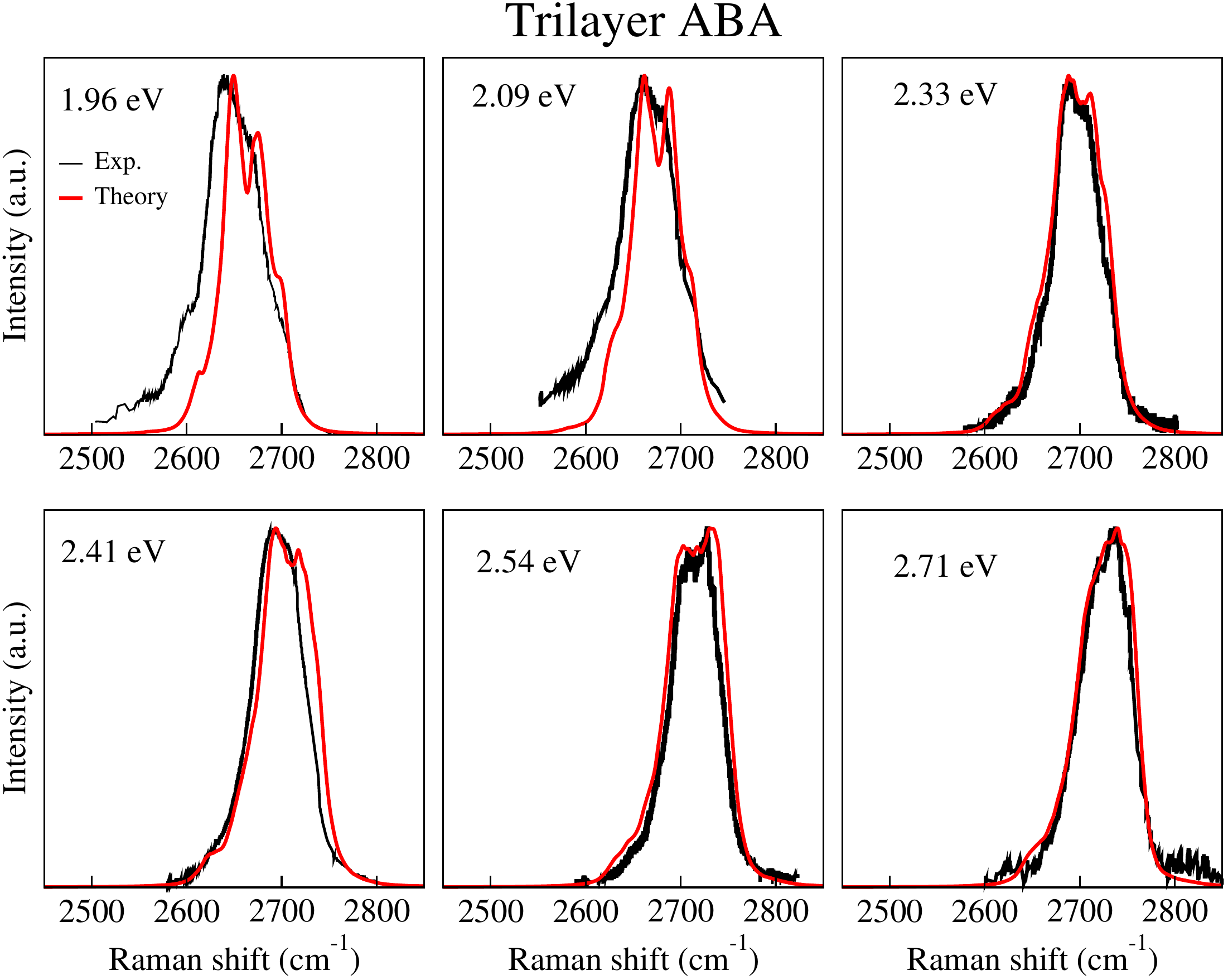}\hspace{0.cm} \hspace{0.2cm}\includegraphics[width=0.49\textwidth]{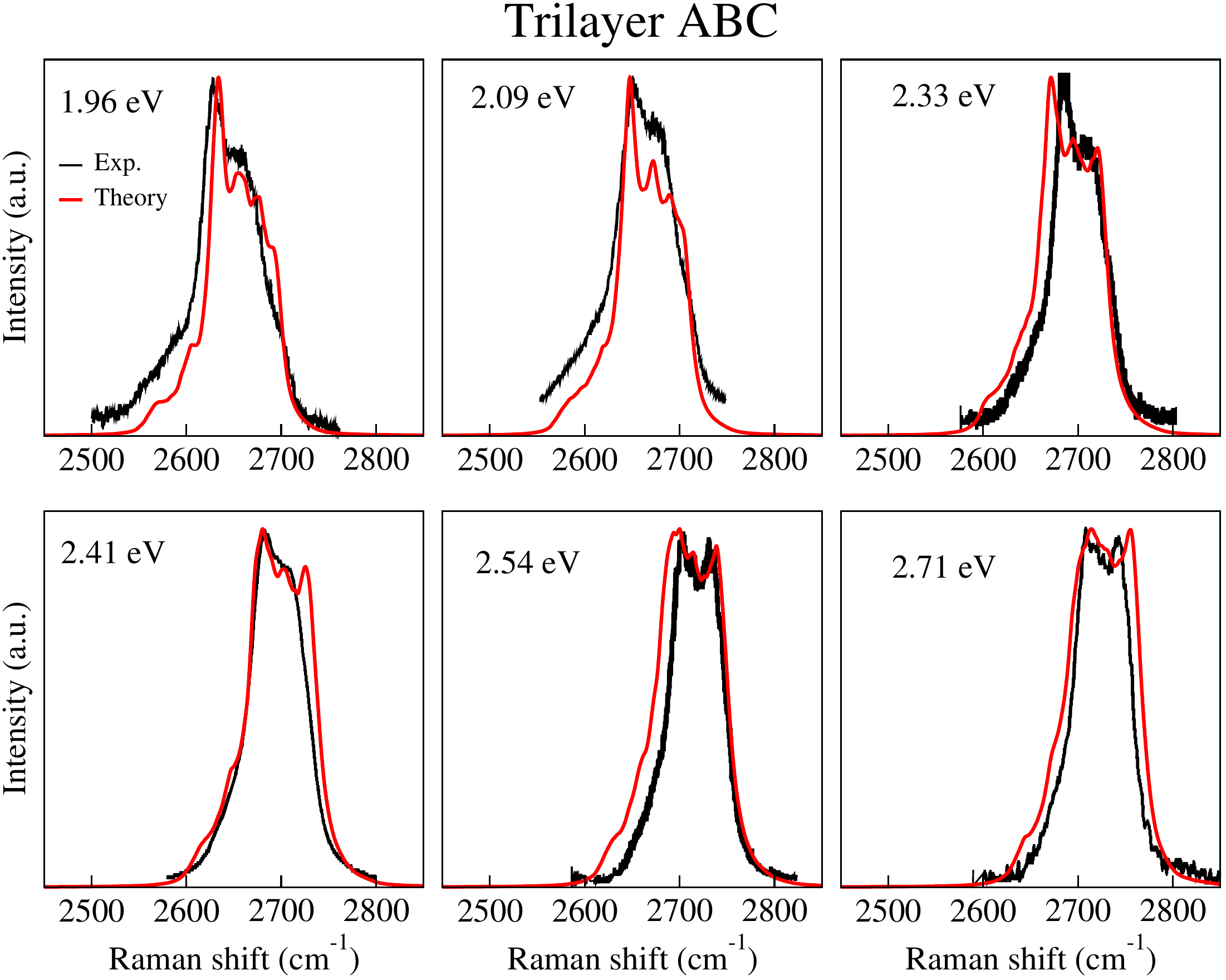}
	\caption{Measured  versus calculated Raman spectra of ABA and 
		ABC trilayer for different laser energies. Experimental data are from Refs.  \cite{cong2011raman,lui2010imaging}.}
	\label{fig_tri}
\end{figure*}
\begin{figure*}[t!]
	\centering 
	\vspace{0.4 cm}
	\includegraphics[width=0.8\textwidth]{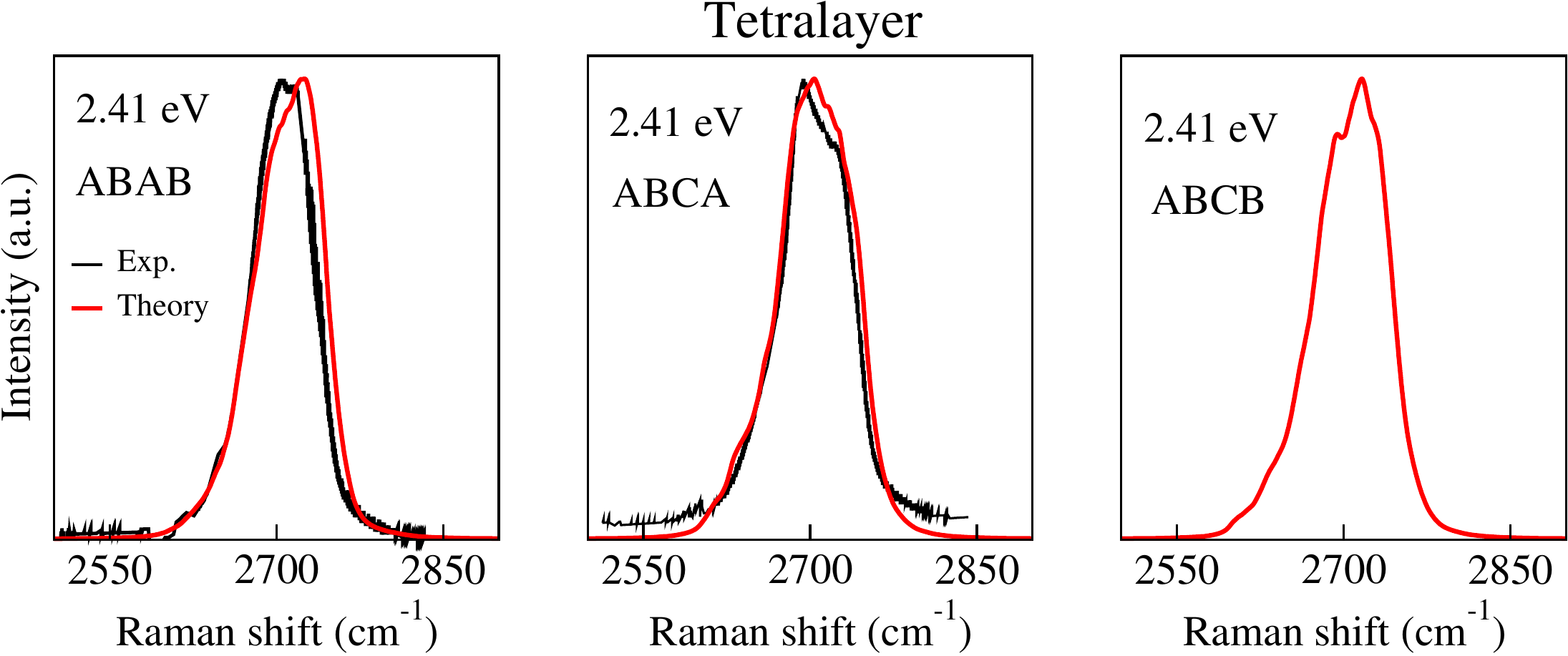}
	\caption{Measured  versus calculated Raman spectra for the three possible stacking in tetralayer graphene at 2.41 eV. Experimental data  are from Ref.   \cite{lui2010imaging}.}
	\label{fig_tetra}
\end{figure*}
\begin{figure*}[t]
	\begin{center}
		\bf{Bulk AB}
	\end{center}
	\includegraphics[width=\textwidth]{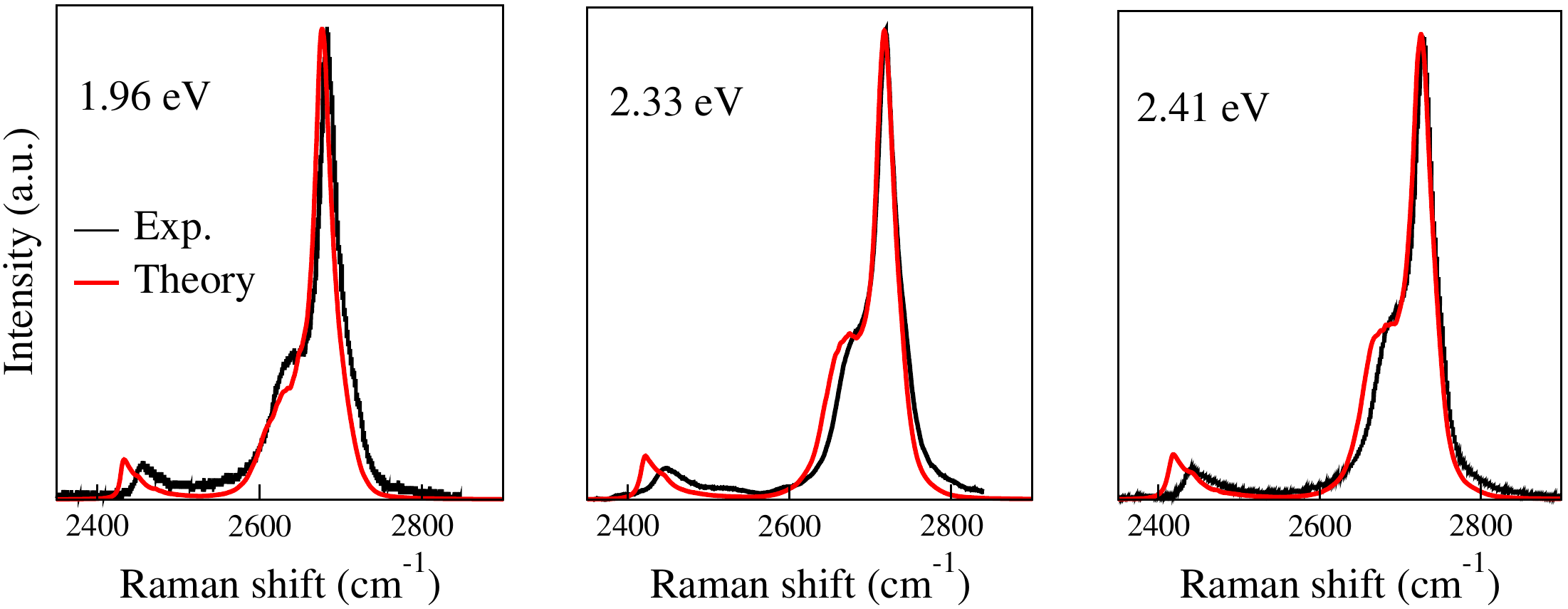} 
	\caption{Comparison between the calculated 2D Raman mode for bulk AB graphite (Bernal graphite) and the experimental Raman spectra obtained from HOPG graphite at different laser energies. Experimental data are from Ref.  \cite{zhang2016review} for 1.96 eV and 2.33 eV and from Ref.  \cite{ferrari2006raman} for 2.41 eV.}
	\label{fig_bernal}
\end{figure*}
\begin{figure*}[t]
	\centering 
	\includegraphics[width=0.85\textwidth]{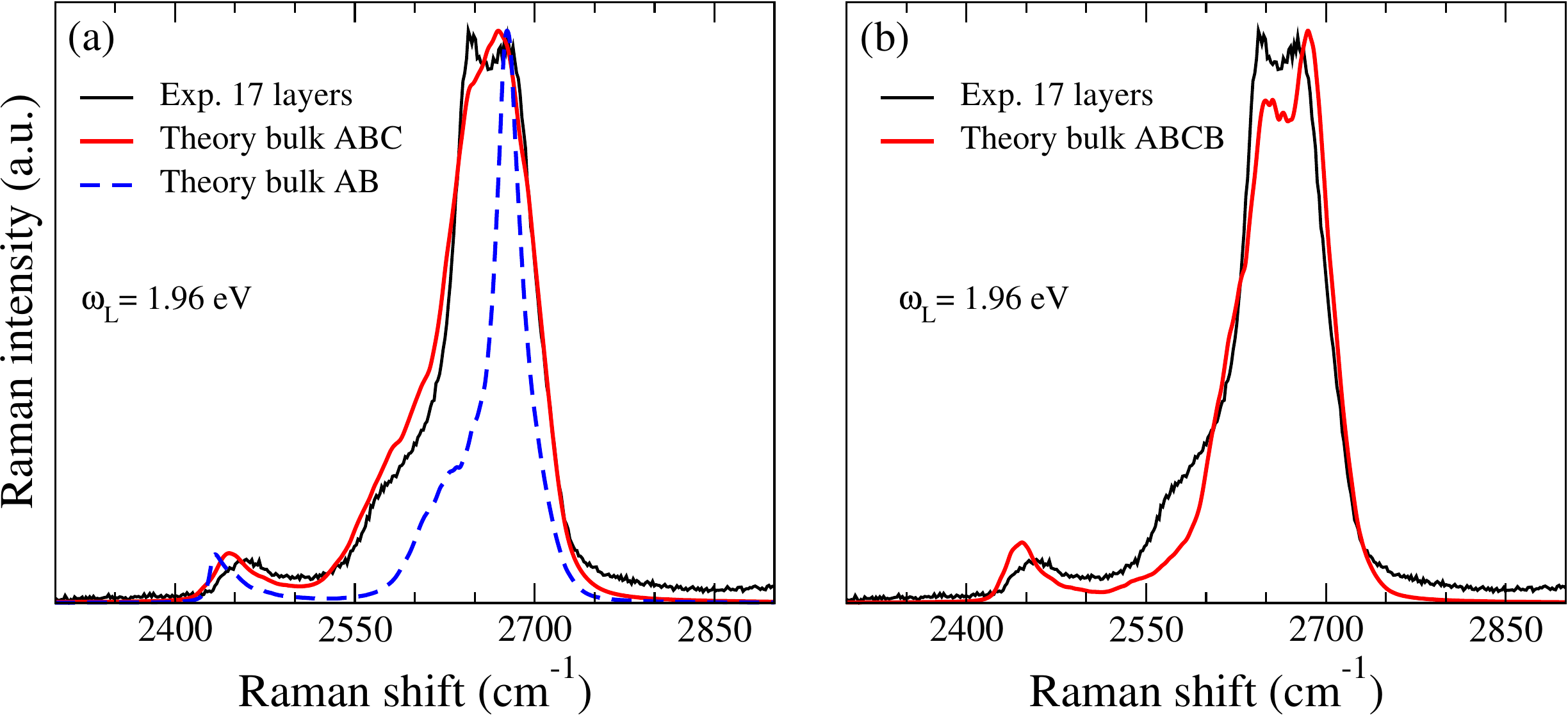}
	\caption{(a) Theoretical spectra for bulk AB, ABC and ABCB stacked graphite against experiments at 1.96 eV. (a) The signature of the long rang ABC stacking at around $\approx 2576$ cm$^{-1}$ is shown. (b) We show the absence of this signature for short rang ABC-stacking in ABCB graphite. The experimental data are from samples composed of approximately $17$ layers of ABC-stacked graphite \cite{henni2016rhombohedral}.}
	\label{fig2}
\end{figure*}
In this work we provide a complete first principles description of the 2D Raman peak in three and four layer graphene for all possible
stackings, as well as for bulk AB, bulk ABC and a periodic
mixing of the two (the so-called ABCB graphite). We present calculations for several laser energies and
we give practical prescriptions to identify long sequences of ABC stacked multilayer graphene. 

Double resonant spectra are calculated from first principles using the method developed in Ref.  \cite{herziger2014two}.
The electrons and phonons bands where first calculated by using the  {\sc Quantum ESPRESSO}\, \cite{QE-2009} code in
the local density approximation with norm-conserving pseudopotentials and an energy cutoff of 70 Ry. 
Electronic integration was performed on k-point grids of $64 \times 64 $ for three and four layer systems and
$64 \times 64 \times 4$ for Bernal graphite. For rhombohedral graphite we use the hexagonal unit cell containing
three layers (6 atoms/unit cell) and a $64 \times 64 \times 4$ k-point grid.  The dynamical matrices and
the electron-phonon coupling were first calculated in linear response on sparse phonon momentum grids ($6\times 6$ for three and four layer graphene and $6\times6\times 3$ for bulk graphites) and then both were Wannier interpolated throghout the Brillouin zone (BZ) using the method of Ref.  \cite{CalandraPhysRevB.82.165111}. 
The electronic bands and phonon frequencies were corrected for the electron-electron interaction as performed in Ref.   \cite{herziger2014two}. 
The double resonant Raman cross section  was calculated on ultra-dense phonon  grids for  reciprocal space integration, namely, grids of of $300\times300$ for 3 and 4 layers and up to $300\times300\times16$ for bulk graphites and electron grids as large as 
$256\times 256$ for few layers and $128\times128\times16$ for bulk graphites. As only a small percentage of the phonon and electron momenta in the grids actually contributes to the cross section, we develop an authomatic technique to identify the subset of  relevant points (see Ref.  \cite{supplemental}).
The electron lifetime was chosen as in Ref.  \cite{herziger2014two} and it was kept the same for all calculations.\\
\begin{figure*}[t]
	\centering 
	\includegraphics[width=0.8\textwidth]{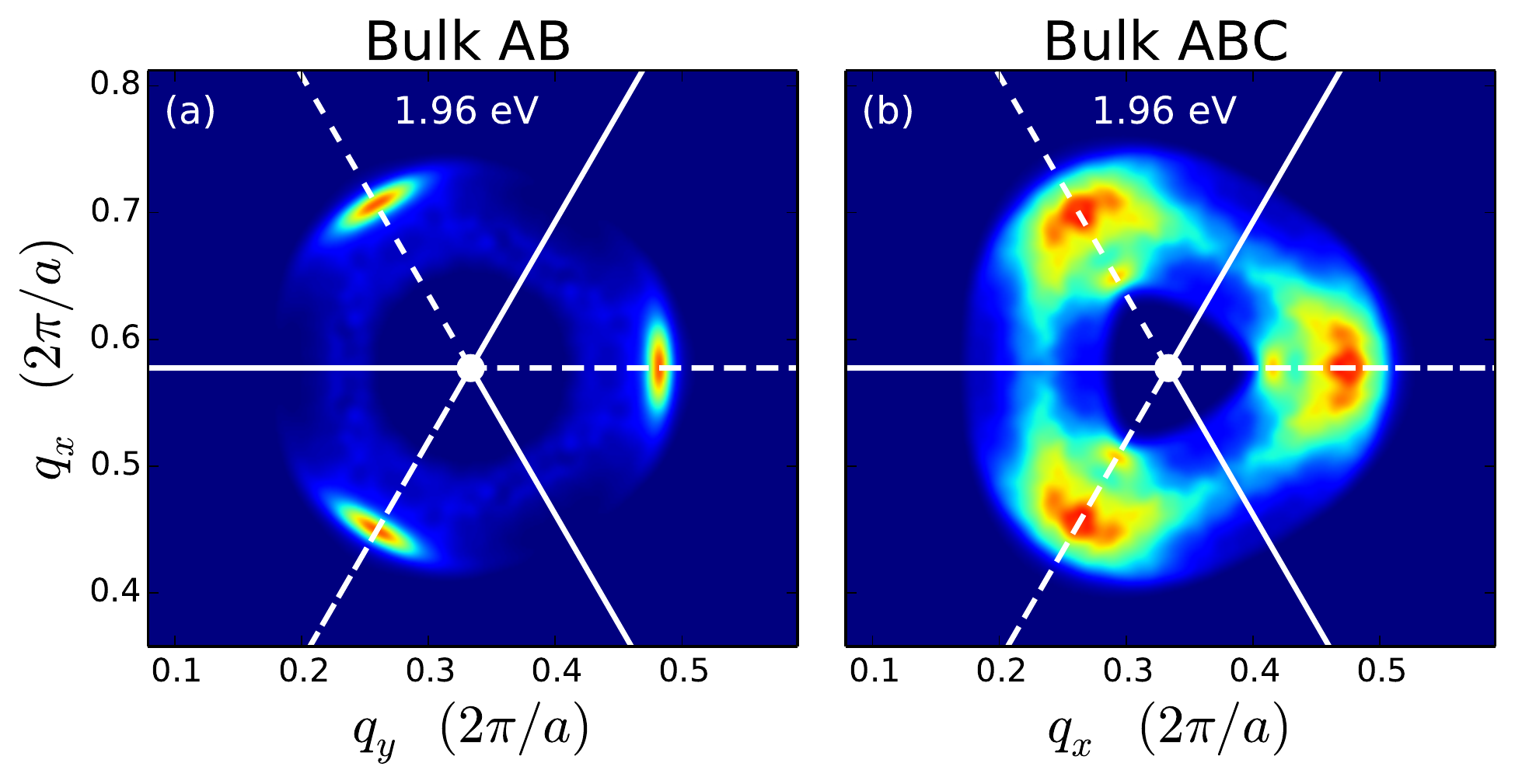}
	\includegraphics[scale = 0.9]{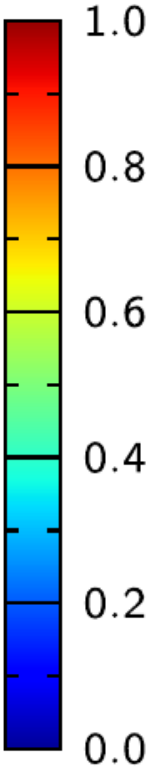}
	\caption{Phonon momenta contributing to the Raman cross 
		section in Bulk AB (a) and Bulk ABC (b) graphite around the $\bf K$ point in the BZ at 1.96 eV. The solid and dashed white lines denote the $\bf K-M$ and $\bf K-\Gamma$
		high-symmetry lines respectively. The color bar indicates the normalized $\bf q$-resolved Raman cross section, where $\bf q$ is the phonon momentum. See supplemental material \cite{supplemental} for relevant definitions.}
	\label{figcontour}
\end{figure*}
We first calculate the double resonant Raman spectra for three and four layers, where a determination of the stacking sequence has been obtained by optical measurements \cite{lui2010imaging}. The results are shown in Fig. \ref{fig_tri}
for three layers and  several laser energies and in
Fig. \ref{fig_tetra} for four layers. Additional results for the
four-layers case are presented in the
supplemental material \cite{supplemental}. 
Overall we find an excellent agreement between our parameter-free {\it ab initio} calculation
and experimental data. We reproduce all 
spectral features in position, width and intensities as well as the laser energy dependence of the spectra. 
Our results clearly demonstrate that the spectra substantially differ from one stacking to the other. 
This is mostly due to the difference in electronic structure between different stackings and to the dominance of symmetric inner processes
(see Fig. \ref{fig1} (b) and supplemental materials  \cite{supplemental}).
Indeed, while along the $\bf \Gamma-K$ direction, the  electronic structure is stacking independent for both three and four layers,
it differs along the $\bf K-M$ high symmetry line. As a consequence, electron-hole pairs are created/destroyed at slightly different points in the BZ 
for the same incident laser energy.  As shown in Fig. \ref{fig1} (b), inner processes  imply electron-hole pairs creation and distraction in the BZ region close to $\bf K-M-K$ and thus the resulting spectrum mostly feel
the difference in electronic structure close to this high symmetry direction. 
A similar effect occurs in four layer graphene (see Fig. \ref{fig_tetra})
Having validated our calculation against experimental data on three
and four layers graphene, we switch to the case of bulk graphite. We first
consider bulk AB graphite (Bernal graphite) for which several experimental data are
available. 
We calculate the spectra for different laser energies
finding an excellent agreement with experimental data (see
Fig. \ref{fig_bernal}). 
We then consider in more details the spectra at $\omega_L=1.96$ eV. 
The 2D peak is composed of a main peak at $\approx 2683$ cm$^{-1}$ and a shoulder around $2640$ cm$^{-1}$, as shown
in Fig. \ref{fig_bernal}. 
Both features are well described by the calculation. The shape and intensities of the D+D$^{''}$ overtone structure at $\approx 2456$ cm$^{-1}$, although at slightly lower energy in the calculation, are also very well reproduced. 

In order to detect signatures of different kinds of stackings, we perform calculations for the case of ABC bulk graphite and ABCB bulk graphite.
ABCB bulk graphite is interesting as it corresponds to a sequence ...[ABC](BAB)[CBA](BCB)... that is an equal mixing of trilayers with rhombohedral (labeled [ABC or CBA]) 
and Bernal (labeled (BAB or BCB))) stackings. Thus, the differences between bulk AB and ABCB stackings can be seen as fingerprint of local rhombohedricity
(i.e. few ABC layers) while the differences between  bulk ABCB  and bulk ABC are signatures of long range rhombohedral order.  
The results are depicted in Fig.\ref{fig2}, where they are compared with the spectrum of Ref.  \cite{henni2016rhombohedral}
that has been tentatively attributed to 17 layers ABC-stacked graphene.  
Both the 2D peaks spectra of bulk ABC and ABCB graphite are substantially broader than the one of Bernal graphite. 
Thus the increased width of the 2D peak at $\omega_L=1.96$ eV is a
fingerprint of short range ABC sequences.

Even if both bulk ABC and bulk ABCB spectra seems similar, they differ for the presence of a feature at $\approx 2560$ cm$^{-1}$ 
in the ABC case that is completely missing in the ABCB stacking. As the bulk ABC structure differs from the bulk ABCB one for the occurrence of long range 
rhombohedral order, the   feature at $\approx 2576$ cm$^{-1}$ at $\omega_L=1.96$ eV  can be seen as a fingerprint of long range 
rhombohedral order. The good agreement between our theoretical calculation and the experimental spectrum in  Ref.  \cite{henni2016rhombohedral}, both from 
what concerns the 2D peak width and shape as well as the presence of the  feature at $\approx 2576$ cm$^{-1}$ suggests that 
the samples in  Ref.   \cite{henni2016rhombohedral} contains long range sequences of rhombohedral stacked multilayer graphene. 

It is worthwhile to discuss a bit more the width of the 2D spectra for
rhombohedral and Bernal graphite as the larger width of the 2D
peak in the former with respect to the latter seems counterintuitive. Indeed, bulk AB-stacked graphite has two couples of $\pi$, $\pi^{*}$ electronic bands in the BZ, while 
ABC graphite only one in the rhombohedral cell. So one could naively think that AB-stacked graphite should have more allowed dipolar transitions. 
However, this is without taking into account the electronic $k_z$
dispersion, that, as shown in supplemental material \cite{supplemental}, is substantially different along K-M-K  in the two case. The different k$_z$ electronic band dispersion implies that different electron and phonon momenta
contribute to the Raman cross section. This is clearly seen in the
in Fig. \ref{figcontour} where the resonant phonon momenta
contributing to the 2D peak cross sections are highlighted in a
contour plot (see also Ref.  \cite{herziger2014two} for more technical explanations). While the Bernal case includes very sharp resonances in phonon
momenta, the resonance is much broader in the rhombohedral case due to
the different band dispersion along $z$. This explain the larger width
of the 2D peak in the rhombohedral case. 
More detailed analysis of electronic transitions contributing to the 2D two-phonon resonant cross section are given in the 
supplemental material \cite{supplemental}. 

In this work we performed parameter-free first principles calculation of the two-phonon resonant 2D and D+D$^{''}$ peaks  in three and four layer graphene for all possible stackings, as well as for bulk AB, ABC and ABCB graphite, that is a  periodic arrangement of AB and ABC graphites. 
Our calculations carried out for several laser energies are in
excellent agreement with experimental data  available for three and
four layers with AB and ABC stacking sequences as well as for Bernal graphite. In the case of ABC-stacked graphite, we compare our
calculation with recently synthesized flakes from Ref.  \cite{henni2016rhombohedral} that were tentatively attributed to 
17 layers ABC sequences. Our theoretical bulk ABC-stacked graphite
spectra confirm this attribution. Furthermore, we have shown how to distinguish between short and long range rhombohedral order.

A. T. is indebted to the IDS-FunMat European network.
M. C. and F. M. acknowledges support from the European Union Horizon 2020 research and  innovation programme under Grant agreement No. 696656-GrapheneCore1, PRACE for awarding us access to resource on Marenostrum at BSC and the computer facilities provided by CINES, IDRIS, and CEA TGCC (Grant EDARI No. 2017091202).
J.-C.C. acknowledges financial support from  the Fédération  Wallonie-Bruxelles through the  ARC entitled 3D Nanoarchitecturing of 2D crystals (N$\degree$ 16/21-077), from the European  Union's Horizon 2020 researchers and innovation programme (N$\degree$ 696656), and from the Belgian FNRS. 

\bibliography{paper}
\bibliographystyle{apsrev4-1}

\newpage



\section*{Supplementary material}

# Supplementary informations For :
# Raman fingerprint of rhombohedral graphite

## I. RAMAN CROSS SECTION AND REDUCTION METHOD

The double resonant Raman intensity for two phonon scattering is given by [1] :

$$I(\omega) = \frac{1}{N_q} \sum_{\mathbf{q},\nu,\mu} I_{\mathbf{q}\nu\mu} \delta(\omega_L - \omega - \omega^{\nu}_{-\mathbf{q}} - \omega^{\mu}_{\mathbf{q}})[n(\omega^{\nu}_{-\mathbf{q}}) + 1][n(\omega^{\mu}_{\mathbf{q}}) + 1], \tag{1}$$

Where $\omega^{\mu}_{\mathbf{q}}$ and $n(\omega^{\nu}_{\mathbf{q}})$ are the phonon frequencies and the Bose distribution for mode $\mu$, respectively. The laser energy is denoted $\omega_L$. The sum is performed over a uniform phonon-momentum grid in the Brillouin zone (BZ) where $N_q$ is the total number of points. The scattering probability for a specific couple of phonon modes $\mu$ , $\nu$ and for a given scattered phonon mometum $\mathbf{q}$ is

$$I_{\mathbf{q}\nu\mu} = \left| \frac{1}{N_k} \sum_{\mathbf{k},\beta} K_{\beta}(\mathbf{k}, \mathbf{q}, \nu, \mu) \right|^2 \tag{2}$$

Where $\mathbf{k}$ here stands for the electron wave vector and $K_{\beta}(\mathbf{k}, \mathbf{q}, \nu, \mu)$, $\beta = 1, ..., 8$ are the scattering probabilities. Explicit expressions for $K_{\beta}(\mathbf{k}, \mathbf{q}, \nu, \mu)$ are given in appendix A of Ref . [1]. The sum in Eq. 2 is performed over a uniform electron-momentum grid in the BZ (called electron grid) where $N_k$ is the total number of points.

Ultra-dense phonons and electrons momentum grids are needed to compute the Raman cross section (summations in Eqs. 1 and 2). Typically, momentum grids of $480 \times 480 \times 1$ were used in Refs. [1, 2] for monolayer and bilayer graphene. However, only a small portion of the grids actually satisfies resonant conditions and contributes to $I(\omega)$. Here below we explain how we menage to condense points in the relevant part of the grids, without calculating the cross section for the portion of the BZ not satisfying the resonant conditon.

### 1. Phonon Grid

For the phonon grid (sum over $\mathbf{q}$ in Eq. 1), we proceed as follows:

1. We first calculate $I(\omega)$ in Eq. 1 for uniform and coarse electron and phonon grids.

2. We calculate the intensity $I(\omega)$ on the coarse phonon grid and select the set of points that gives a substantial contribution to the cross section, namely those satisfying the condition:

$$I_{\mathbf{q}\mu\nu} > \frac{I_{max}}{S_{ph}} \tag{3}$$

Where $S_{ph}$ is a parameter (typical values ranges from 10 to 200). This subset of q-points constitutes the so-called resonant region for phonon momenta. We also verify that Eq. 1 restricted to the resonant region leads to the same $I(\omega)$ for the coarse grid.

### 2. Electron grid

A similar procedure is adopted for identifying the resonant region in for electron-momenta. To do that, we have to examine the exact form of the kernel $K_{\beta}(\mathbf{k}, \mathbf{q}, \nu, \mu)$ (appendix A of Ref. [1]). The denominators in the expression giving $K_{\beta}(\mathbf{k}, \mathbf{q}, \nu, \mu)$ are of the following kind (for the sake of simplicity we neglect the phonon energy):

$$(\omega_L - \epsilon^{\pi^*}_{\mathbf{k}} + \epsilon^{\pi}_{\mathbf{k}} - i\gamma_{\mathbf{k}}) \times (\omega_L - \epsilon^{\pi^*}_{\mathbf{k}\pm\mathbf{q}} + \epsilon^{\pi}_{\mathbf{k}} - i\gamma_{\mathbf{k}}) \times (\omega_L - \epsilon^{\pi^*}_{\mathbf{k}+\mathbf{q}} + \epsilon^{\pi}_{\mathbf{k}\pm\mathbf{q}} - i\gamma_{\mathbf{k}}) \tag{4}$$

Where $\epsilon^{\pi^*}_{\mathbf{k}\pm\mathbf{q}}$ ($\epsilon^{\pi}_{\mathbf{k}\pm\mathbf{q}}$) are the energies of the electronic $\pi^*$ and $\pi$ states, $\gamma_{\mathbf{k}}$ is the electronic lifetime and $\omega_L$ is the laser energy. According to the double resonant model [1, 3] the first term (creation of the electron-hole pair :

$\omega_L - \epsilon_{\mathbf{k}}^{\pi^*} + \epsilon_{\mathbf{k}}^{\pi} - i\gamma_{\mathbf{k}})$ and the third term (annihilation of the electron-hole pair : $\omega_L - \epsilon_{\mathbf{k}\pm\mathbf{q}}^{\pi^*} + \epsilon_{\mathbf{k}\pm\mathbf{q}}^{\pi} - i\gamma_{\mathbf{k}})$ in Eq. 4 are responsible for the double resonance during the Raman scattering [1]. The non-resonant regions for electron-momenta are the set of $k-$points of the grid where at least one of these two denominators is off-resonance. In order to avoid these regions we adopted the following procedure :

1. We first calculate $I(\omega)$ in Eq. 1 for a coarse electron grid.

2. We determine the resonant region for electrons from the following conditions that have to be simultaneously verified for each phonon momentum $\mathbf{q}$ determined in the previous subsection:

$$| \epsilon_{\mathbf{k}}^{\pi^*} - \epsilon_{\mathbf{k}}^{\pi} | \quad < \omega_L + S_{el}\gamma \tag{5}$$

$$| \epsilon_{\mathbf{k}\pm\mathbf{q}}^{\pi^*} - \epsilon_{\mathbf{k}\pm\mathbf{q}}^{\pi} | \quad < \omega_L + S_{el}\gamma \tag{6}$$

Where $S_{el}$ is a parameter ranging tipically from 5 to 15.

3. We recalculate the $I(\omega)$ and verify that it does not differ significantly from the one calculated with the full coarse grid.

Once the resonant regions of the electron's and phonon's grid are obtained on coarse grids, we proceed as follows. Fist, we converge the sum over $\mathbf{k}$ in Eq. 2 keeping the coarse phonon grid, namely we consider the electron-momenta belonging to the resonant region and we shift them randomly of an amount given by the distance between the $k-$points times a random number between 0 and 1. This generates many randomly displaced resonant regions. We calculate the cross section summing over all electron grids and calculate the spectrum until we are at convergence.

Second, we generate a new randomly displaced resonant region for phonon momenta as we did for the electrons. For this new grid we converge again the sum over electron-momenta as done previously. We then calculate the cross section for each randomly displaced resonant region for phonon momenta. If n-randomly displaced resonant regions are generated, we calculate the cross section and obtain the n-th approximant $I^n(\omega)$. We then calculate

$$\overline{I}^n(\omega) = \frac{1}{n}\sum_1^n I^i(\omega) \tag{7}$$

When $\overline{I}^n(\omega)$ is converged, we stop.

For the case of graphene, our procedure allows to gain a factor of 300 on the computational effort with the respect to the brute force calculation (no use of resonant regions). This factor is less for bilayer, trilayer and tetralayer graphene. This is expected since the resonance in the case of graphene is very sharp (a sharp 2D peak) so that the resonant regions (for electrons and phonons) in the BZ in graphene are sharper then the case of bilayer, trilayer and tetralayer graphene.

## II. GW CORRECTION FOR ELECTRONS AND PHONONS

Phonon dispersions are obtained from ab-initio density functional theory (DFT) calculations in the linear-response scheme using LDA as exchange correlation kernel. The phonon frequencies were corrected with GW, following the work done in Refs. [1, 4, 5] for single-layer graphene. More specifically, we correct the TO branches dispersion in order to reproduce Kohn anomaly [6] near the K point in the BZ, namely we define

$$(\omega_{q\nu}^{GW})^2 = \frac{1}{2}erfc\left(\frac{|\mathbf{q} - \mathbf{K}_\alpha|\frac{a_0}{2\pi} - 0.2}{0.05}\right) \times [(\alpha_k - 1) \times (\omega_{q\nu}^{LDA})^2 + \triangle] \tag{8}$$

Here $erfc$ is the error function, $\nu$ is a label for the TO phonon modes, $a_0$ is the graphene lattice constant ($a_0 = 2.4595$ Å ) and $K_\alpha$ is the closest vector to $q$ among those equivalent to K. The constants $\alpha_K$ and $\triangle$ are defined as $\alpha_K = 1.61$ and $\triangle = 42.195$ Ryd$^2$ and are identical for all the TO phonon modes.

The electronic structure (eigenvalues $\epsilon_{\mathbf{k}}^i$ and eigenvectors $|\mathbf{k}, i\rangle$) is first obtained from ab-initio calculation using LDA as exchange correlation kernel and interpolated using Wannier function. The Fermi velocity is increased by +18% in order to reproduce the GW bands structure that is in a good agreement with angle-resolved photo-emission spectra (ARPES) measurements in graphite [1, 7]. Namely ($\epsilon_f$ is the Fermi energy)

$$\epsilon_k^{i,GW} = (\epsilon_k^i - \epsilon_f) \times 1.18 + \epsilon_f \tag{9}$$

Figs. 1,2,3,4, 5 show the LDA+GW electronic bands and phonon frequencies compared to LDA calculation around the **K** point in the BZ for trilayer, tetralayer and bulk AB-stacked graphite (Bernal graphite) respectively. The LDA+GW phonon frequencies show an enhanced Kohn anomaly, as expected. Furthermore, the TO splitting near the K point is larger after the GW correction than in LDA. For trilayer graphene, the maximum splitting of the TO phonons near the K point is 6 cm⁻¹ in LDA, while in GW it is 10 cm⁻¹. In the case of Bernal graphite, while the TO branches are almost degenerate within LDA,a splitting of 12 cm⁻¹ is observed near the K point after including the GW correction. The TO splitting is of great importance in the calculation of the Raman cross section since it determines the position of the sub-contributions of different symmetry-allowed processes to the 2D mode, as it will be discussed in the analysis section of this letter.

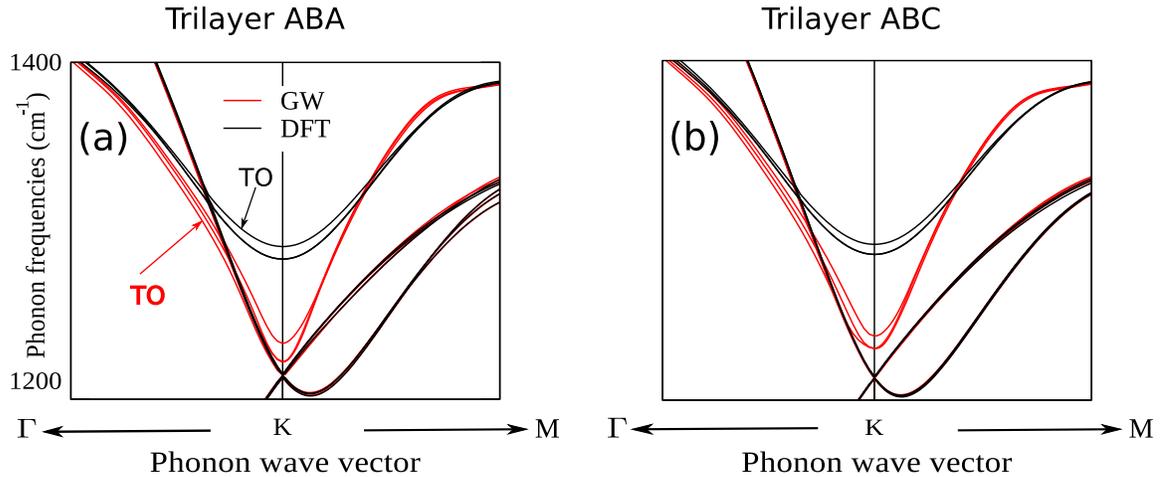

Figure 1: Phonon frequencies in the LDA and LDA+GW approximations for (a) ABA and (b) ABC trilayer graphene around the **K** point in the BZ.

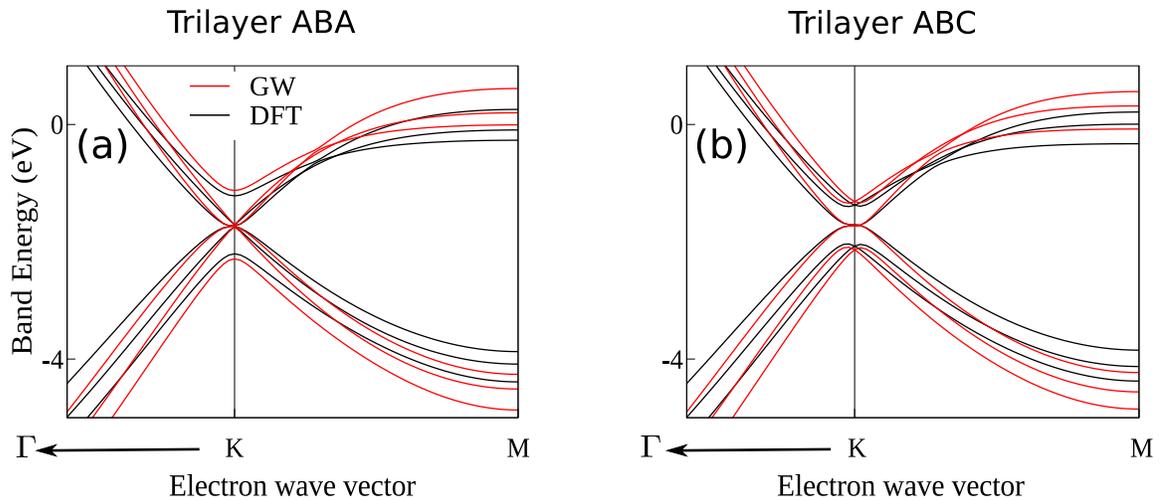

Figure 2: Electronic structure in the LDA and LDA+GW approximation for (a) ABA and (b) ABC trilayer graphene around the **K** point in the BZ.

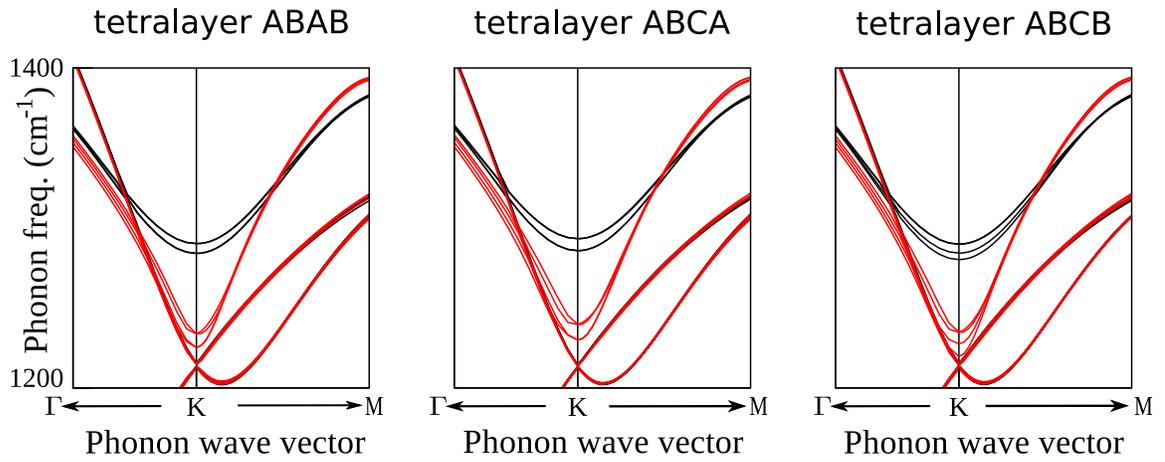

Figure 3: Comparison between LDA and LDA+GW-corrected phonon dispersion for the three polytypes of tetralayer graphene, namely : ABAB, ABCA and ABCB stacking around the **K** point in the BZ.

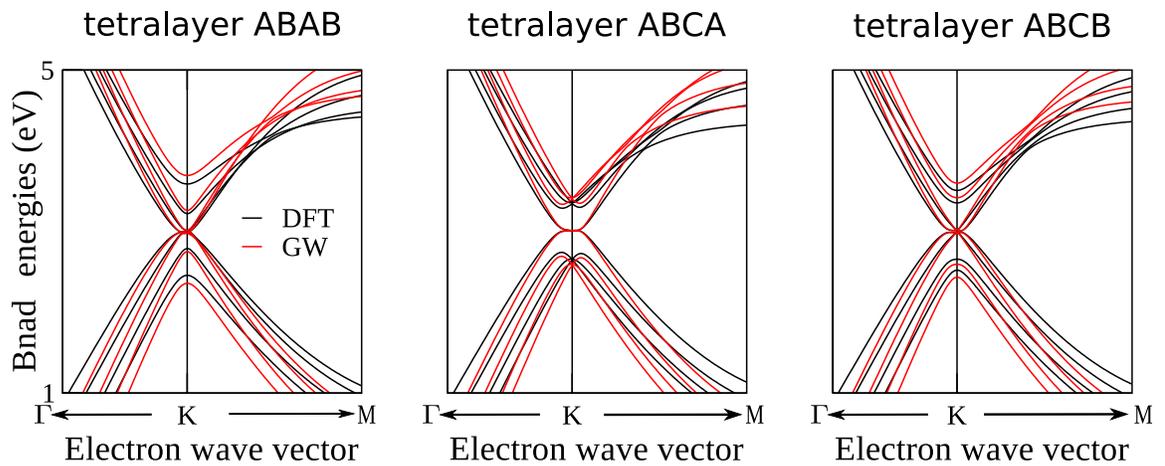

Figure 4: Comparison between LDA and LDA+GW-corrected electronic bands dispersion for the three polytypes of tetralayer graphene, namely : ABAB, ABCA and ABCB stacking around the **K** point in the BZ.

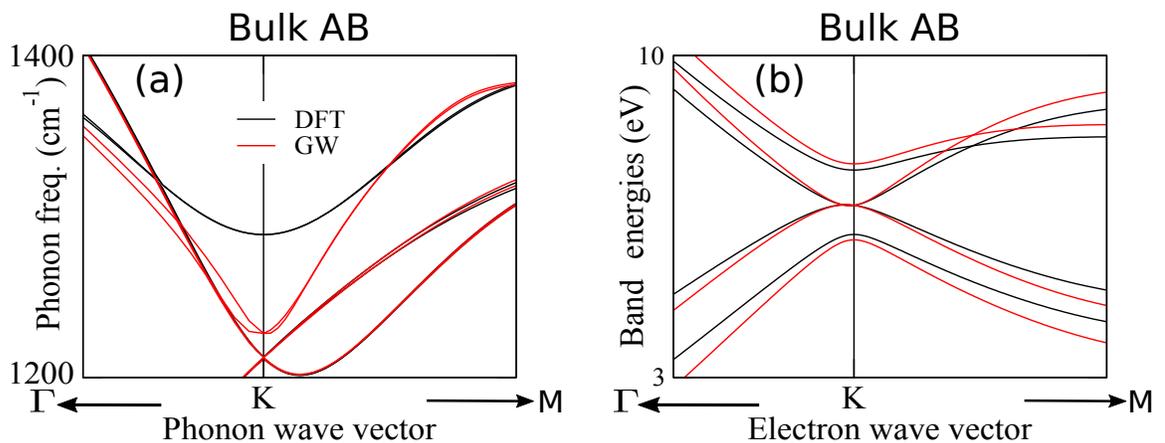

Figure 5: Comparison between LDA and LDA+GW corrected phonon dispersion (left panel) and electronic bands (right panel) for Bernal graphite around the **K** point in the BZ.

**Tetralayer graphene**

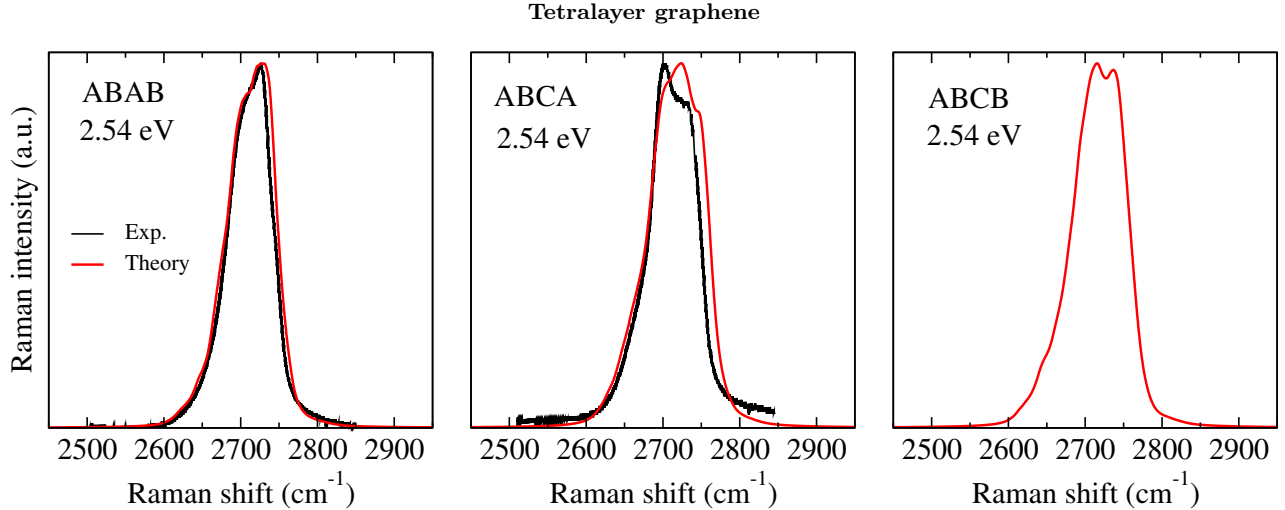

Figure 6: Measured versus calculated Raman spectra for the three possible stackings in tetralayer graphene at 2.54 eV. Experimental data are from Ref. [8].

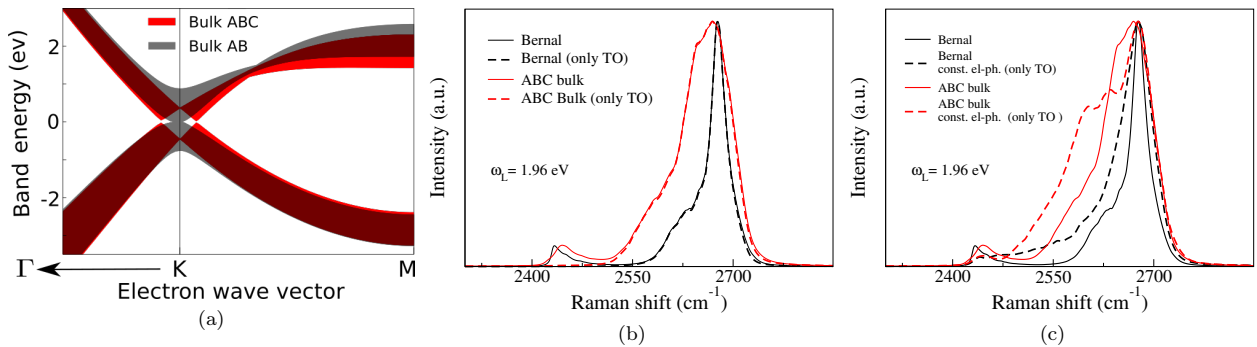

Figure 7: (a) Bernal (bulk AB) versus rhombohedral (bulk ABC) electronic bands dispersion. All bands with for all possible values of $k_z$ are represented in the plane $k_z = 0$. This explain the thickness of the band structures in (a). (b) Contribution of the traversal optical (TO) modes (dashed lines) to the 2D Raman mode for bulk AB and bulk ABC graphites (solid lines) (c) The same as in panel (b) except that the TO contribution is now calculated with constant electron-phonon matrix elements.

### III. TETRALAYER GRAPHENE

Fig. 6 show the calculated spectra for all possible alternate stackings in tetralayer graphene compared with the experimental results at 2.54 eV. The agreement is excellent as it was the case for 2.41 eV excitation energy (see main text of this letter).

### IV. WIDTH OF THE 2D PEAK IN BULK GRAPHITES

The larger width of the 2D peak in bulk ABC graphite with respect to the bulk AB graphite seems counterintuitive as explained in the main text of this letter. Indeed, AB-stacked graphite has two couples of $\pi$, $\pi^*$ (4 atoms/unit cell) electronic bands in the BZ, which gives rise to four allowed Raman precesses while ABC-stacked graphite has only one couple of $\pi$, $\pi^*$ (2 atoms/unit cell) and one Raman process. So one could naively think that AB-stacked graphite should have more allowed dipolar transitions and,consequently, the 2D peak should be broader. However, this is without taking into account the electronic $\mathbf{k_z}$ dispersion that, as shown in Fig. 7a, is substantially different along $\mathbf{K} - \mathbf{M} - \mathbf{K}$ in the two case. The different $\mathbf{k_z}$ electronic band dispersion implies that different electron and

phonon momenta contribute to the Raman cross section.

In order to explicitly verify the effect of the electronic band dispersion on the 2D peak, we calculate the full Raman cross section with two approximations. First, we restrict the calculation to scattering with transverse optical phonons only (both for electrons and holes), as shown in Fig. 7b. This approximation leads to indistinguishable spectra for the 2D peak from the full calculation, meaning that the 2D feature is only due to transverse optical (TO) modes. On the contrary, under this approximation the D+D″ peak disappears as it is in part due to scattering to longitudinal acoustic phonons, as it was shown in the case of graphene [1]. Second, we calculate the spectrum restricted to TO modes and we set the electron-phonon matrix elements of these phonons to unity, regardless of the electronic band index. The results of this approximation are depicted in Fig. 7c. As it can be seen, for both AB and ABC-stacked graphite, all the peaks occurring in the complete calculation are still present. However the intensities are different. Under this last approximation the intensities are simply related to the number of optically-active dipolar-processes contributing to the Raman cross section as they are not anymore weighted by the electron-phonon matrix element. As a result of its different $k_z$ dispersion, rhombohedral graphite has a substantially larger number of optically active dipolar processes and this explains the origin of its broader spectrum despite the reduced number of bands in the BZ.

## V. SPECTRA ANALYSIS

The Raman process responsible for the 2D peak in few layer graphene and graphite can be described as a 4 steps scattering process [1] (see Fig. 1 (b) in the main text of this letter). First, an electron-hole pair is created due to annihilation of the incoming photon. Second, the electron (or the hole) is scattered by a phonon with momentum $\mathbf{q}$. Third, the electron (or the hole) is scattered by a phonon with momentum $-\mathbf{q}$ ($-\mathbf{q}$ from momentum conservation). Finally, the electron-hole pair recombines and the outgoing photon is emitted. According to this description, many processes can take place, depending on the electronic states between which the creation (recombination) or the scattering by the phonon occurs. However, dipole selection rules, and phonons symmetries, reduce the total number of possible processes to a set of symmetry allowed processes, for example, Graphene has only one process, bilayer four and trilayer 15 (For a review see Ref. [9]). In this section we present several types of analysis that were performed on the obtained spectra for trilayer graphene and bulk AB-stacked graphite in order to understand the structure of the 2D.

### A. Trilayer graphene

#### Polarization study

Figs. 8 (a) show, for ABA and ABC trilayer graphene, light polarization analysis of the Raman cross section at 1.96 eV (the same analysis at 2.71 eV is presented in Figs. 9 (a)). We define the xx (resp. xy) contribution as the contribution where the incident light is polarized along the x direction and the scattered light is polarized along the x (resp. y) direction. The contributions yy and yx are defined in the same way. The symmetry $x \equiv y$ is preserved in our calculation. Furthermore, if we define $I_{\parallel}$ to be the parallel contribution to the Raman cross section where the scattered light is parallel to the incident one and, conversely, the perpendicular contribution $I_{\perp}$, to be the contribution where the scattered light is perpendicular to the incident one, then trilayer graphene have the same ratio $\frac{I_{\parallel}}{I_{\perp}} \approx 2.7$ as for graphene [1]. This ratio is found to be constant and independent of the laser excitation energy.

#### Inner-outer decomposition

Figs. 8 (b) show, for ABA and ABC trilayer graphene, the electron-hole ($e-h$) contribution to the Raman cross section for both stackings as well as the inner-outer decomposition at 1.96 eV (the same analysis at 2.71 eV is presented in Figs. 9 (b)). Definitions of the $e-h$ (also hole-hole: $h-h$ and electron-electron : $e-e$) processes can be found in App. A of Ref. [1]. A process is called inner if the resonant phonon wave vector stems from a sector of $\pm 30°$ next to the $\mathbf{K} - \mathbf{\Gamma}$ direction with respect to $\mathbf{K}$. Conversely, outer processes have phonon wave vectors from $\pm 30°$ next to the $\mathbf{K} - \mathbf{M}$ directions. As in single layer [1] and bilayer graphene [2], the $e-h$ processes are dominant compared to $e-e$ and $h-h$ ones. Furthermore we notice that the Inner contribution is dominant over the outer one for $\epsilon_L = 1.96$ eV. By increasing the laser energy, the outer contribution increases and becomes almost comparable to the inner one for laser energy $\epsilon_L = 2.71$ eV.

*Processes decomposition*

In Figs. 8 (c) we show the decomposition over the 15 allowed processes of the $2D$ mode in trilayer graphene for the ABA and ABC stacking. These processes are labeled following the electron transitions between filled and empty states during the Raman scattering. Namely, the process $P_i^{jl}$ refers to a scattering where the electron is initially at the filled state $i$, it will be excited to the empty state $j$ and then scatter by a phonon to the empty state $l$. We notice that the interference between these processes is totally additive and the their sum gives the $e - h$ scattering spectrum. This additive interference plays an important role in the 2D structure as it enhances the sum of the sub-contributions and leads to the experimental observed shape of the 2D peak. Although there are theoretically 15 allowed processes for trilayer graphene, we observe in Figs. 8 (c), for ABA and ABC trilayers, that 6 of them have a negligible contribution (the processes that are not present in these figures have negligible contribution to the 2D peak). Furthermore, we noticed that some of the remaining 9 processes give the almost the same contribution and thus are called degenerate processes, namely, processes of the from: $P_i^{jl}$ and $P_k^{lj}$ are degenerate. These are the 6 following processes (degenerate by pair) : $(P_2^{56}, P_1^{65})$, $(P_3^{46}, P_1^{64})$ and $(P_2^{45}, P_2^{54})$. The remaining 3 processes, namely, the processes: $P_1^{55}$, $P_3^{44}$ and the $P_1^{66}$ give independent contributions. In view of the previous observations, one can say that theoretically, the 2D peak in ABA abd ABC trilayer graphene is composed by 6 different contributions or sub-peaks. This was already observed experimentally in Refs. [8, 10] where the obtained 2D peaks are shown to be reproducible using at least 6 Gaussian functions fitting. However the others of Refs. [8, 10] could not explain why only 6 contributions are observed even if symmetry-allowed processes are 15 in total. Here we give the theoretical explanation of this fact.

*Symmetric-antisymmetric decomposition*

In trilayer graphene the symmetry of electronic bands are as follow [9] (for bands indexing see Fig. 1 (b) in the main paper) :

$$\pi_1 \rightarrow T^- \quad \pi_2 \rightarrow T^+ \quad \pi_3 \rightarrow T^- \qquad \text{and} \qquad \pi_1^* \rightarrow T^- \quad \pi_2^* \rightarrow T^+ \quad \pi_3^* \rightarrow T^-$$

A Raman process $P_i^{jl}$ is said to be symmetric if the electronic scattering with a phonon occurs between the electronic states $\{j, l\}$ that have the same symmetry ($T^- \rightarrow T^-$ or $T^+ \rightarrow T^+$). Conversely, a Raman process is said to be antisymmetric if the scattering occurs between electronic states that have different symmetry ($T^- \rightarrow T^+$ or $T^+ \rightarrow T^-$). Figs. 8 (d) show, for ABA and ABC trilayer, the decomposition of the 2D peak upon symmetric and antisymmetric processes. We notice that the symmetric contribution is dominant over the antisymmetric one, however, the antisymmetric contribution increases with laser energy (see Fig. 9 for analysis at 2.71 eV).

**q***-resolved Raman cross section (contour plot)*

In Fig. 10 we present the contour plot of the Raman cross section for ABA abd ABC trilayer graphene at different excitation energies. Contour plots show the dependence of the Raman cross section on the phonon wave vector $\mathbf{q}(q_x, q_y)$ in the BZ, more precisely, in Fig. 10 we plot the function $I(q_x, q_y)$ defined by :

$$I(q_x, q_y) = \sum_{\mu, \nu}^{N_m} I_{\mathbf{q}\nu\mu} \tag{10}$$

where $I_{\mathbf{q}\nu\mu}$ is defined in Eq. 2 and $N_m$ is the number of modes ($N_m = 3 \times N_{atoms}$ where $N_{atoms}$ is the number of atoms in the unit cell). This plot allows us to visualize the resonant regions in the BZ.

## B. Bulk AB-graphene

The same analysis (as discussed in the previous four paragraphs) was performed for AB-stacked graphite (Bernal graphite) and the results are depicted in Fig. 11. The most important thing to notice in the case Bernal graphite concerns processes decomposition presented in Fig. 11 (c). Bernal graphite have 4 symmetry allowed processes that are indicated by four indexes. In Fig. 11 (c) a $P_{ij}^{lm}$ processes refers to a Raman scattering where an electron is initially at the filled state $i$, it will be excited to the empty state $j$ and then scattered by a phonon to the empty state $l$. on the other hand, The hole is scattered by a phonon from the state $i$ to the state $j$. Finally the recombination process

takes place between the $l$ and $j$ states. Based on the processes analysis presented in Fig. 11 (c), we can say that the main peak of the 2D structure (the peak around $\sim 2700$ cm$^{-1}$ at 1.96 eV) is mainly due to the process $P_{22}^{33}$, while the other processes, namely, $P_{11}^{44}$, $P_{13}^{42}$ and $P_{24}^{31}$ form the shoulder of the 2D structure. If we combine this results with the inner outer decomposition from Fig. 11 (b) and the symmetric-antisymmetric decomposition from Fig. 11 (d), we can say, in a more detailed way, that the main peak of the 2D structure in Bernal graphite is mainly due to the inner contribution of the symmetric process $P_{22}^{33}$. The outer-antisymmetric contribution forms the shoulder of the 2D structure.

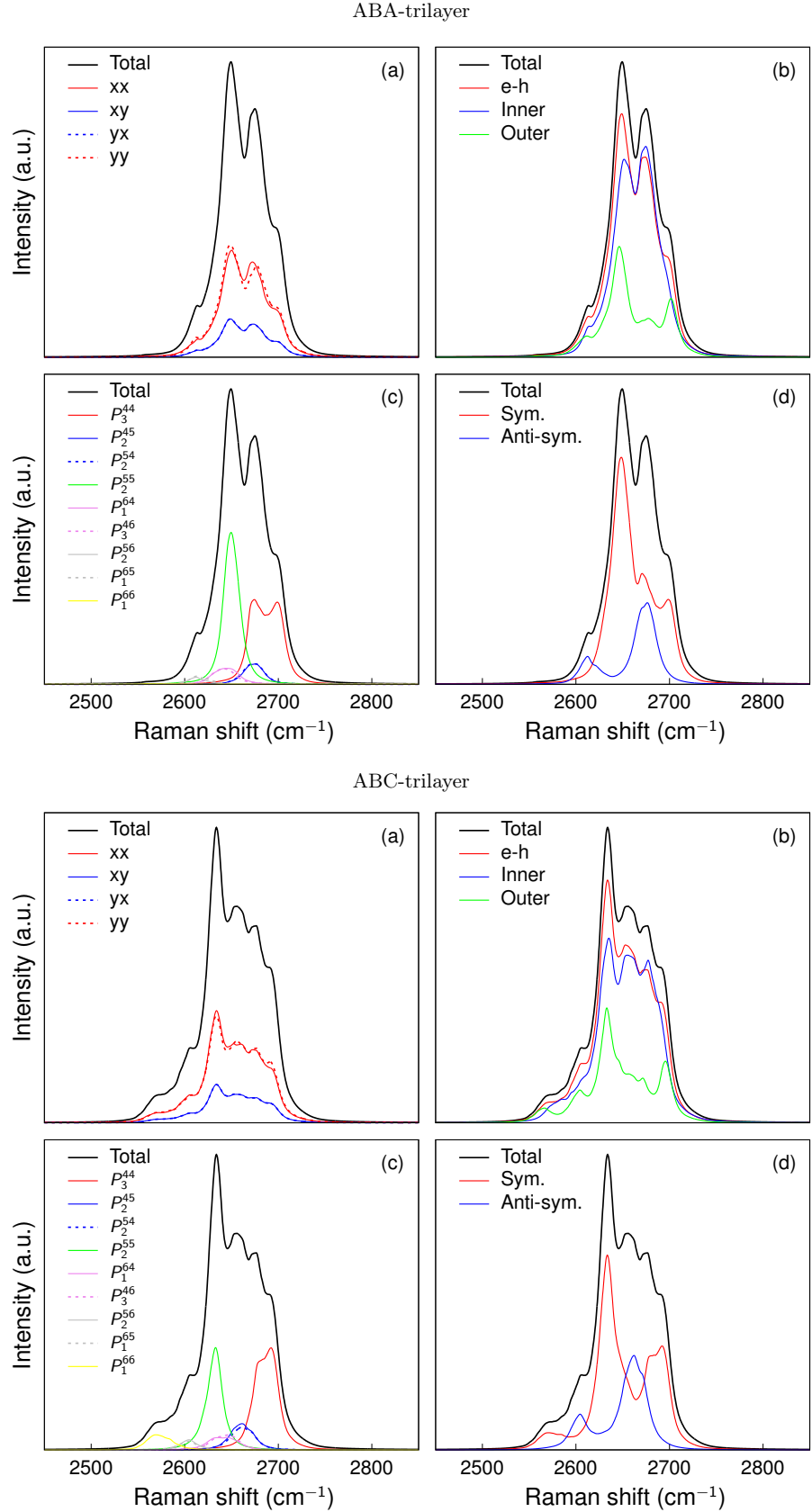

Figure 8: Analysis of the 2D Raman mode in ABA (upper panel) and ABC (lower panel) trilayer graphene at 1.96 eV. (a) polarization analysis, (b) $e-h$ contribution and inner outer decomposition, (c) Process analysis: dashed and solid lines with the same color indicate degenerate processes. (d) symmetric and anti-symmetric decomposition. See Sec. V for relevant definitions

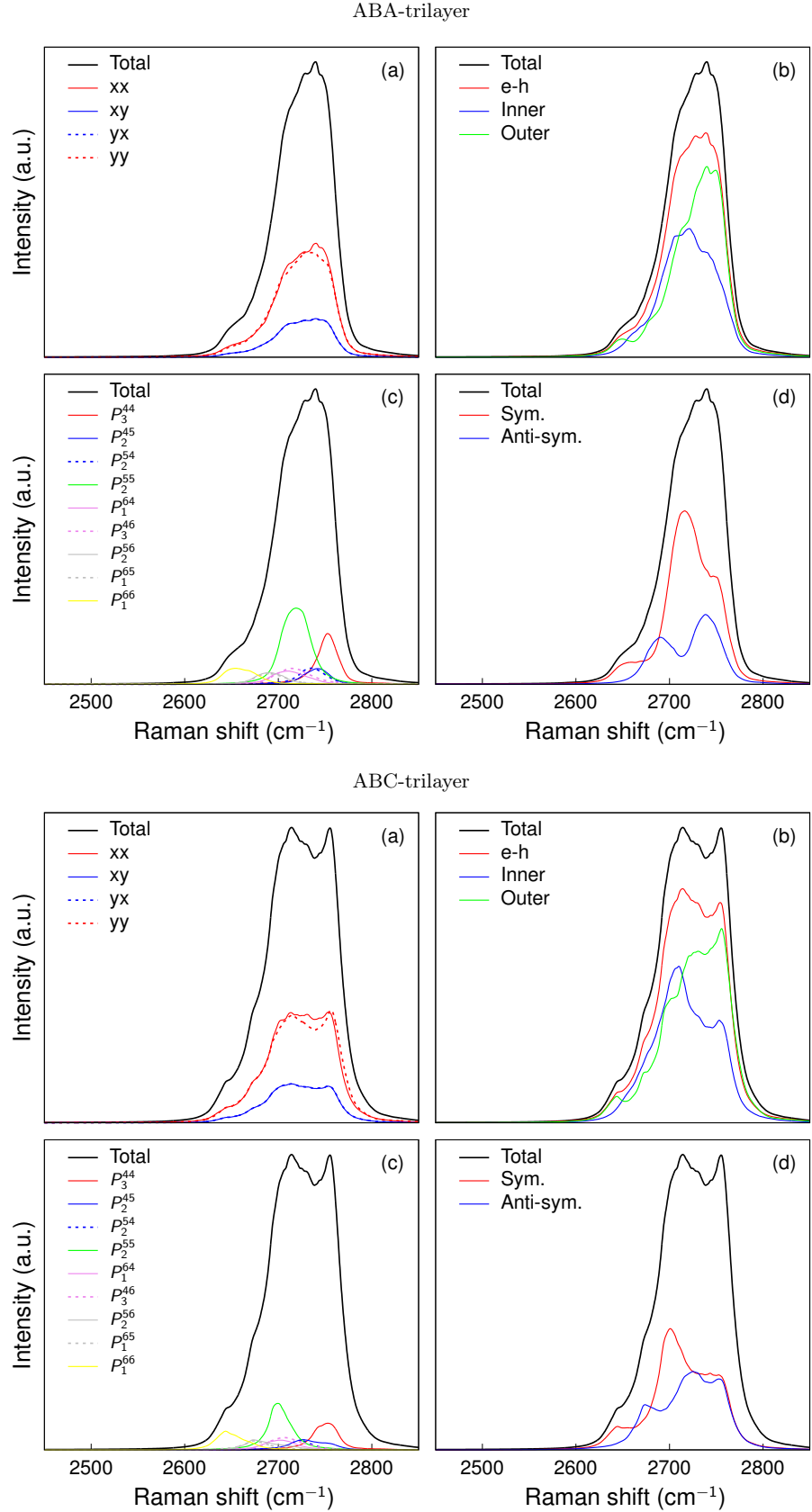

Figure 9: Analysis of the 2*D* Raman mode in ABA (upper panel) and ABC (lower panel) trilayer graphene at 2.71 eV. (a) polarization analysis, (b) $e - h$ contribution and inner outer decomposition, (c) Process analysis (d) symmetric and anti-symmetric decomposition. See Sec. V for relevant definitions

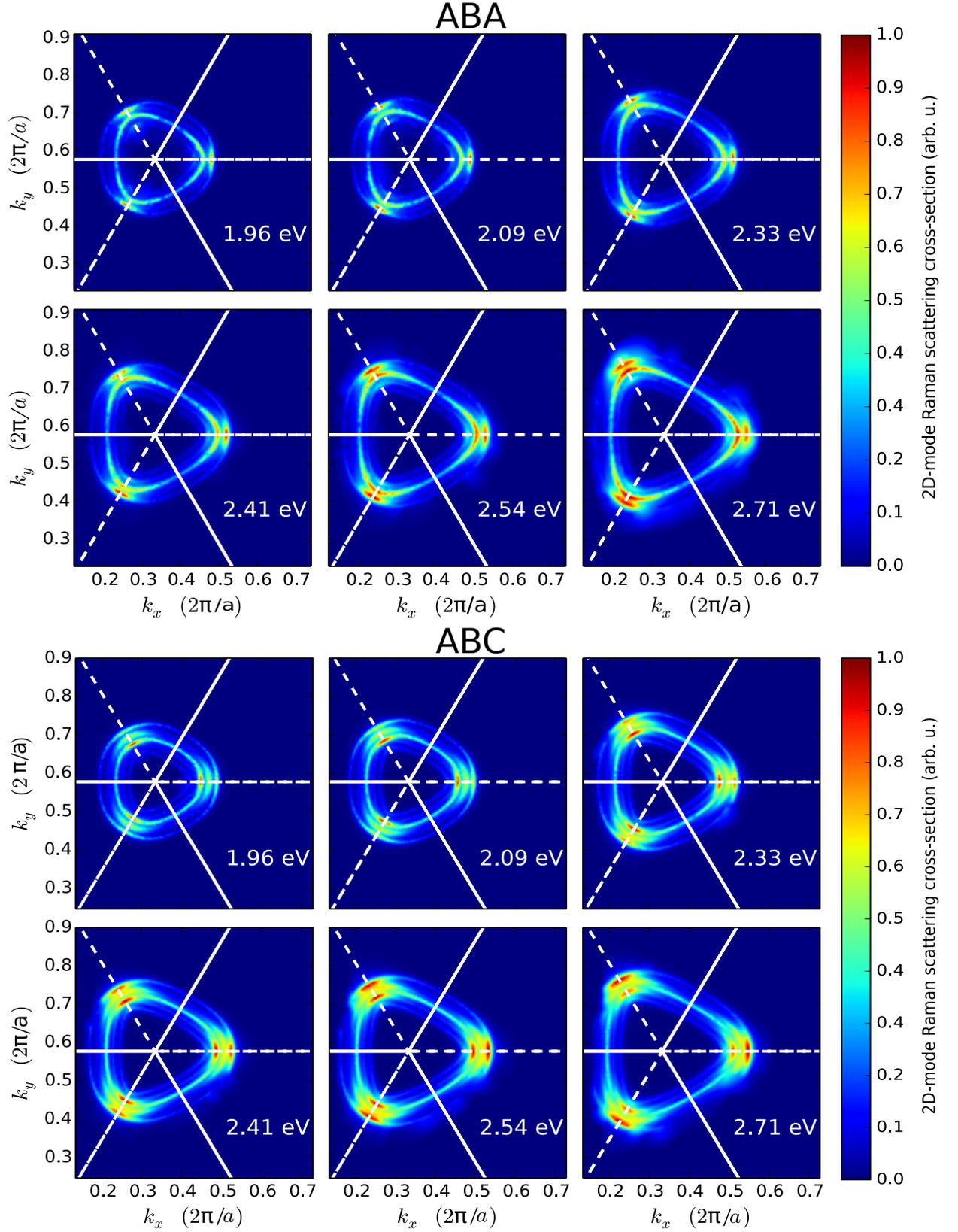

Figure 10: Contour plot of the Raman intensity for ABA (upper panel) and ABC (lower panel) trilayer graphene for different excitation energies. The dashed lines indicates the inner directions corresponding to scattering between Dirac points across the M point, while filled lines indicates the outer directions corresponding to scattering across the $\Gamma$ point in the BZ. See Sec. V for relevant definitions

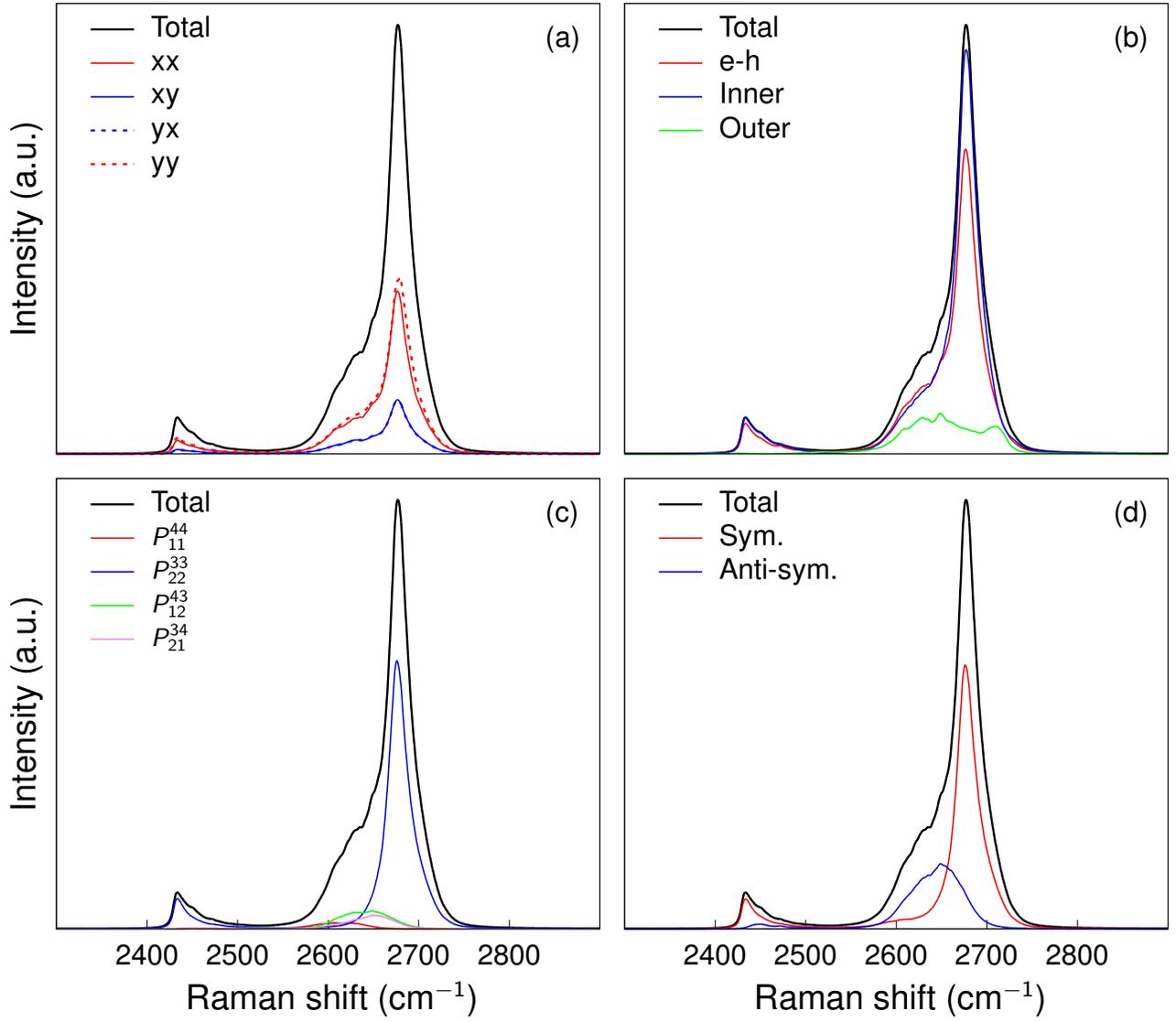

Figure 11: Analysis of the $2D$ Raman mode of bulk AB-stacked graphite (Bernal graphite) at 1.96 eV. (a) polarization analysis, (b) $e - h$ contribution and inner outer decomposition, (c) Process analysis (d) symmetric and anti-symmetric decomposition. See Sec. V for relevant definitions


\end{document}